\documentclass[conference]{IEEEtran}
\IEEEoverridecommandlockouts

\usepackage{cite}
\usepackage{amsmath,amssymb,amsfonts}
\usepackage{algorithmic}
\usepackage{graphicx}
\usepackage{textcomp}
\usepackage{xcolor}
\usepackage{booktabs}
\usepackage{multirow}
\usepackage{hyperref}
\usepackage{algorithm}
\usepackage{subcaption}

\hypersetup{
    colorlinks=true,
    linkcolor=blue,
    citecolor=blue,
    urlcolor=blue
}

\def\BibTeX{{\rm B\kern-.05em{\sc i\kern-.025em b}\kern-.08em
    T\kern-.1667em\lower.7ex\hbox{E}\kern-.125emX}}

\begin{document}

\title{Predictive-LoRA: A Proactive and Fragmentation-Aware Serverless Inference System for LLMs}

\author{
\IEEEauthorblockN{
Yinan Ni\textsuperscript{1},
Xiao Yang\textsuperscript{2},
Yuqi Tang\textsuperscript{3},
Zhimin Qiu\textsuperscript{4},
Chen Wang\textsuperscript{5},
Tingzhou Yuan\textsuperscript{6}
}
\IEEEauthorblockA{
\textsuperscript{1}University of Illinois at Urbana--Champaign, Urbana, IL, USA\\
\textsuperscript{2}Santa Clara University, Santa Clara, CA, USA\\
\textsuperscript{3}New York University, New York, NY, USA\\
\textsuperscript{4}University of Southern California, Los Angeles, CA, USA\\
\textsuperscript{5}University of Missouri--Kansas City, Kansas City, MO, USA\\
\textsuperscript{6}Boston University, Boston, MA, USA
}
}

\maketitle

\begin{abstract}
The serverless computing paradigm offers compelling advantages for deploying Large Language Model (LLM) inference services, including elastic scaling and pay-per-use billing. However, serving multiple fine-tuned LLMs via Low-Rank Adaptation (LoRA) in serverless environments faces critical challenges: reactive adapter loading causes significant cold start latency, and frequent adapter swapping leads to severe GPU memory fragmentation. In this paper, we present Predictive-LoRA (P-LoRA), a proactive and fragmentation-aware serverless inference system for LoRA-based LLMs. P-LoRA introduces two key innovations: (1) a lightweight LSTM-based traffic predictor that forecasts adapter demand and proactively prefetches hot adapters from host memory to GPU, reducing cold start latency by up to 68\%; and (2) a page-based adapter memory management mechanism inspired by operating system virtual memory, which keeps GPU memory utilization above 87\% even under heterogeneous adapter ranks. We evaluate P-LoRA using production-like workloads derived from the Azure Functions trace. Experimental results demonstrate that P-LoRA achieves 1.52$\times$ higher throughput than S-LoRA while reducing the average Time-To-First-Token (TTFT) by 35\% under high concurrency scenarios.
\end{abstract}


\section{Introduction}
\label{sec:intro}

Large Language Models (LLMs) have demonstrated remarkable capabilities across diverse natural language processing tasks~\cite{brown2020language}. The ``pretrain-then-finetune'' paradigm has become the standard approach for adapting foundation models to specific domains~\cite{hu2022lora}. Low-Rank Adaptation (LoRA) enables parameter-efficient fine-tuning by injecting trainable low-rank decomposition matrices into transformer layers, reducing trainable parameters by orders of magnitude while maintaining competitive performance.

The serverless computing paradigm presents an attractive deployment model for LLM inference services. By abstracting infrastructure management and offering pay-per-use billing, serverless platforms can efficiently handle the bursty and unpredictable nature of LLM workloads~\cite{shahrad2020serverless}. However, deploying LoRA-based LLMs in serverless environments faces two fundamental challenges.

First, existing systems such as S-LoRA~\cite{sheng2024slora} adopt a \textit{reactive} approach to adapter management, loading adapters from host memory to GPU only when requests arrive. This introduces substantial cold start latency, particularly during traffic bursts when multiple adapters must be loaded simultaneously. Our profiling reveals that adapter loading accounts for up to 35\% of end-to-end latency under high concurrency.

Second, LoRA adapters vary significantly in size depending on their rank configuration. Frequent loading and unloading of heterogeneous adapters causes severe GPU memory fragmentation. While vLLM's PagedAttention~\cite{kwon2023vllm} effectively manages KV cache fragmentation, adapter weights present distinct challenges due to their static nature within an inference session and varying sizes across adapters.

In this paper, we present Predictive-LoRA (P-LoRA), a serverless inference system that addresses both challenges through proactive prediction and fragmentation-aware memory management. Our key insight is that adapter access patterns in production workloads exhibit temporal locality and predictability~\cite{shahrad2020serverless}, which can be exploited for proactive prefetching.

P-LoRA makes the following contributions:

\begin{itemize}
\item We design a lightweight LSTM-based traffic predictor that forecasts adapter demand using historical access patterns. The predictor achieves an average 86\% accuracy (up to 89\% under optimal settings) with only 2.3ms inference overhead, enabling proactive adapter prefetching.

\item We propose a page-based adapter memory management mechanism that partitions adapter weights into fixed-size pages, eliminating external fragmentation and enabling flexible memory allocation for heterogeneous adapter ranks.

\item We implement P-LoRA as an extension to vLLM and evaluate it using production-derived workloads from the Azure Functions trace. P-LoRA achieves 1.52$\times$ throughput improvement over S-LoRA while reducing average TTFT by 35\%.
\end{itemize}

\section{Background and Motivation}
\label{sec:background}

\subsection{LoRA and Multi-Adapter Serving}

Low-Rank Adaptation~\cite{hu2022lora} adapts pretrained models by adding low-rank decomposition matrices to specific weight matrices. For a pretrained weight matrix $\mathbf{W}_0 \in \mathbb{R}^{d \times k}$, LoRA introduces:
\begin{equation}
\mathbf{W} = \mathbf{W}_0 + \Delta\mathbf{W} = \mathbf{W}_0 + \mathbf{B}\mathbf{A}
\label{eq:lora}
\end{equation}
where $\mathbf{B} \in \mathbb{R}^{d \times r}$ and $\mathbf{A} \in \mathbb{R}^{r \times k}$, with rank $r \ll \min(d, k)$. This reduces trainable parameters from $d \times k$ to $r \times (d + k)$.

Multi-adapter serving systems like S-LoRA~\cite{sheng2024slora} and Punica~\cite{chen2023punica} enable concurrent serving of multiple LoRA adapters on shared GPUs. They maintain the base model in GPU memory and dynamically load adapters as needed. However, these systems react to requests rather than anticipating them, leading to cold start overhead.

\subsection{Serverless LLM Inference Challenges}

Recent work on serverless LLM inference~\cite{fu2024serverlessllm} has highlighted the cold start problem, where model checkpoint loading dominates startup latency. ServerlessLLM addresses this through multi-tier checkpoint loading but focuses on base model loading rather than adapter management.

We identify two key challenges specific to serverless multi-LoRA serving:

 When a request arrives for an adapter not currently in GPU memory, the system must load it from host memory. For a rank-64 LoRA adapter on Llama2-7B, this takes approximately 28-45ms depending on PCIe bandwidth and concurrent memory transfers.

Different adapters have different memory footprints based on their rank. A rank-8 adapter requires roughly 13MB, while rank-64 requires 100MB+ for Llama2-7B. Continuous loading and unloading creates fragmented memory regions that cannot accommodate new adapters despite sufficient total free memory.

\subsection{Workload Predictability}

Analysis of production serverless workloads~\cite{shahrad2020serverless} reveals significant temporal patterns. We observe that adapter access follows diurnal patterns with recurring hot adapters during specific time windows. This predictability motivates our proactive approach.

\section{System Design}
\label{sec:design}

\begin{figure}[t]
\centering
\includegraphics[width=0.85\columnwidth]{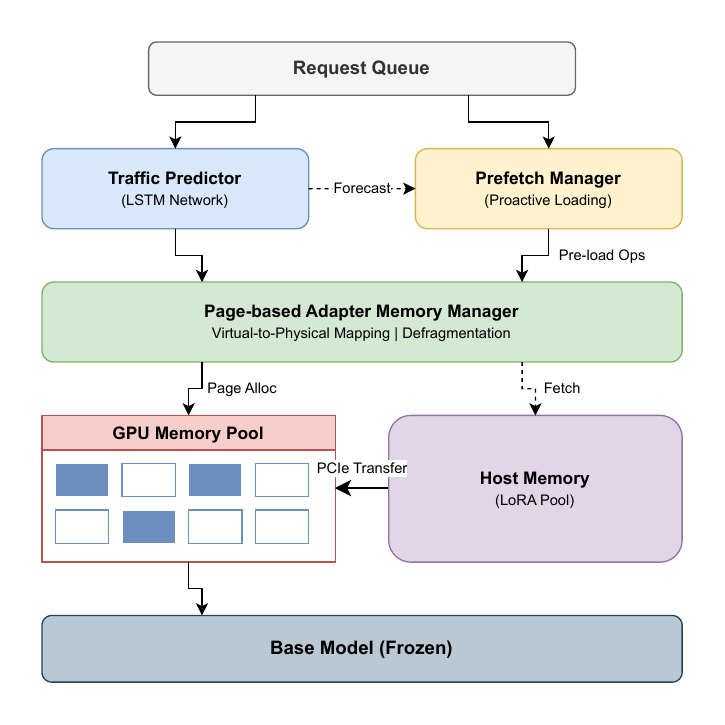}
\caption{P-LoRA system architecture. The traffic predictor analyzes request patterns to forecast adapter demand. The prefetch manager proactively loads hot adapters to GPU memory. The page-based memory manager eliminates fragmentation through virtual-to-physical mapping.}
\label{fig:architecture}
\end{figure}

P-LoRA consists of three main components: (1) a traffic predictor that forecasts adapter demand, (2) a prefetch manager that proactively loads predicted adapters, and (3) a page-based memory manager that handles adapter allocation. Figure~\ref{fig:architecture} illustrates the system architecture.

\subsection{Traffic Prediction}
\label{sec:prediction}

We employ a lightweight LSTM network to predict adapter access patterns. The predictor operates on a sliding window of historical access counts, producing demand forecasts for the next prediction interval.

\textbf{Input Representation.} For each adapter $i$, we maintain a time series of access counts $\{c_i^{t-w}, c_i^{t-w+1}, ..., c_i^{t-1}\}$ over a window of $w$ intervals. We normalize counts and encode adapter identity through learned embeddings.

\textbf{Model Architecture.} We use a 2-layer LSTM with 64 hidden units followed by a fully connected layer that outputs predicted access probability for each adapter. The model is trained using Adam optimizer with learning rate $10^{-3}$ and batch size 64 to minimize cross-entropy loss:
\begin{equation}
\mathcal{L} = -\sum_{i=1}^{N} y_i \log(\hat{p}_i) + (1-y_i)\log(1-\hat{p}_i)
\label{eq:loss}
\end{equation}
where $y_i$ indicates whether adapter $i$ is accessed in the next interval and $\hat{p}_i$ is the predicted probability.

\textbf{Online Learning.} The predictor continuously updates its parameters using recent observations, adapting to workload shifts. We employ a replay buffer of the most recent 10,000 observations and perform incremental updates every 100 requests.

\subsection{Proactive Prefetching}
\label{sec:prefetch}

The prefetch manager translates predictions into loading decisions. We define a prefetch threshold $\theta$ and load adapters with predicted probability $\hat{p}_i > \theta$ that are not currently in GPU memory.

\textbf{Prefetch Scheduling.} To avoid interfering with active inference, prefetching is scheduled during idle periods between batches. We implement double buffering where adapters are loaded into a staging area before being promoted to the active pool.

\textbf{Eviction Policy.} When GPU memory is constrained, we evict adapters using a combined recency-frequency-prediction score:
\begin{equation}
s_i = \alpha \cdot \text{LRU}_i + \beta \cdot \text{freq}_i + \gamma \cdot \hat{p}_i
\label{eq:eviction}
\end{equation}
where $\text{LRU}_i$ is the recency score, $\text{freq}_i$ is the access frequency, and weights $\alpha, \beta, \gamma$ are tuned empirically.

\subsection{Page-Based Memory Management}
\label{sec:memory}

Inspired by virtual memory paging in operating systems~\cite{kwon2023vllm}, we partition adapter weights into fixed-size pages to eliminate external fragmentation.

\textbf{Page Structure.} We define a page size $P$ (default 2MB) and store adapter weights across multiple pages. For an adapter with weight size $S$, we allocate $\lceil S/P \rceil$ pages. Pages are non-contiguous in physical memory but present a contiguous virtual view to the inference engine.

\textbf{Page Table.} Each adapter maintains a page table mapping logical page indices to physical page locations. During inference, we translate virtual addresses to physical addresses using this mapping. The translation overhead is negligible as it occurs once per batch.

\textbf{Defragmentation.} Background compaction periodically coalesces free pages, though page-based allocation significantly reduces the need for compaction compared to block-based approaches.

\begin{figure}[t]
\centering
\includegraphics[width=0.9\columnwidth]{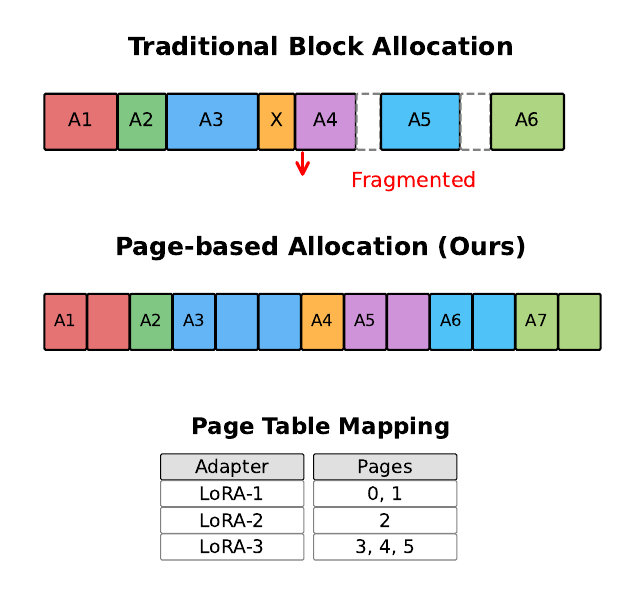}
\caption{Comparison of traditional block allocation (top) versus page-based allocation (bottom). Block allocation leads to fragmentation as adapters of varying sizes are loaded and evicted. Page-based allocation maintains uniform pages with virtual mapping, eliminating external fragmentation.}
\label{fig:memory}
\end{figure}

Figure~\ref{fig:memory} illustrates the difference between traditional block allocation and our page-based approach.

\section{Implementation}
\label{sec:implementation}

We implement P-LoRA as an extension to vLLM~\cite{kwon2023vllm}, adding approximately 3,200 lines of Python and 800 lines of CUDA code.

\textbf{Predictor Integration.} The LSTM predictor runs asynchronously on CPU, processing access logs and generating predictions every 100ms. Predictions are communicated to the GPU scheduler via shared memory to minimize overhead.

\textbf{Memory Manager.} We extend vLLM's block manager to support page-based adapter allocation. The page table is maintained in pinned CPU memory for fast GPU access. We implement custom CUDA kernels for scatter-gather operations that read adapter weights from non-contiguous pages.

\textbf{Batch Scheduling.} We modify the scheduler to incorporate prefetching decisions. When forming a batch, the scheduler checks prediction results and initiates prefetch operations for anticipated adapters during the current batch's execution.

\section{Evaluation}
\label{sec:evaluation}

\subsection{Experimental Setup}

\textbf{Hardware.} We conduct experiments on a server with 8$\times$ NVIDIA A100 (40GB) GPUs, 512GB DDR4 memory, and dual AMD EPYC 7763 processors. Unless otherwise specified, we report single-GPU results.

\textbf{Models.} We evaluate on Llama2-7B, Llama2-13B~\cite{touvron2023llama}, and Mistral-7B~\cite{jiang2023mistral}. We generate synthetic LoRA adapters with ranks $r \in \{8, 16, 32, 64\}$ applied to query and value projection matrices.

\textbf{Workloads.} We derive workload traces from the Azure Functions dataset~\cite{shahrad2020serverless}, mapping function invocations to adapter requests where each unique function identifier corresponds to a distinct LoRA adapter. We scale inter-arrival times to achieve target request rates from 10 to 500 requests/second. Input/output lengths are sampled from the ShareGPT dataset~\cite{zheng2023judging}, which contains real-world conversation distributions.

\textbf{Baselines.} We compare against:
\begin{itemize}
\item \textbf{vLLM}~\cite{kwon2023vllm}: State-of-the-art LLM serving with merged LoRA adapters.
\item \textbf{S-LoRA}~\cite{sheng2024slora}: Scalable multi-LoRA serving with unified paging for KV cache.
\item \textbf{dLoRA}~\cite{wu2024dlora}: Dynamic orchestration of LoRA requests and adapters.
\end{itemize}

\textbf{Metrics.} We measure throughput (requests/second), Time-To-First-Token (TTFT), Time-Per-Output-Token (TPOT), memory utilization, and fragmentation ratio.

\subsection{Cold Start Latency}

\begin{figure*}[t]
\centering
\includegraphics[width=0.95\textwidth]{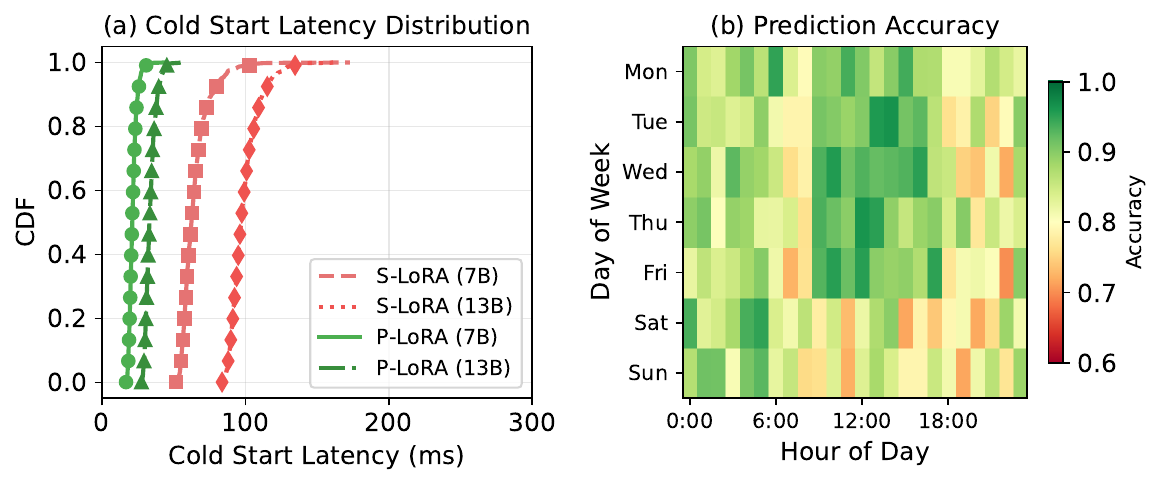}
\caption{(a) CDF of cold start latency for different systems and model sizes. P-LoRA significantly reduces cold start latency through proactive prefetching. (b) Prediction accuracy heatmap across different times, showing higher accuracy during regular business hours.}
\label{fig:cold_start}
\end{figure*}

Figure~\ref{fig:cold_start} shows the cold start latency distribution. P-LoRA's proactive prefetching significantly reduces cold start occurrences. For Llama2-7B, the median cold start latency decreases from 68ms (S-LoRA) to 22ms (P-LoRA), a 68\% reduction. The improvement stems from prefetching anticipated adapters before requests arrive.

The prediction accuracy varies with time of day and day of week. Business hours on weekdays show higher accuracy (88-92\%) due to more regular access patterns, while weekend evenings show lower accuracy (70-78\%) due to sporadic usage. The overall average accuracy across all time periods is 86\%.

\subsection{Throughput and Latency}

\begin{figure*}[t]
\centering
\includegraphics[width=0.95\textwidth]{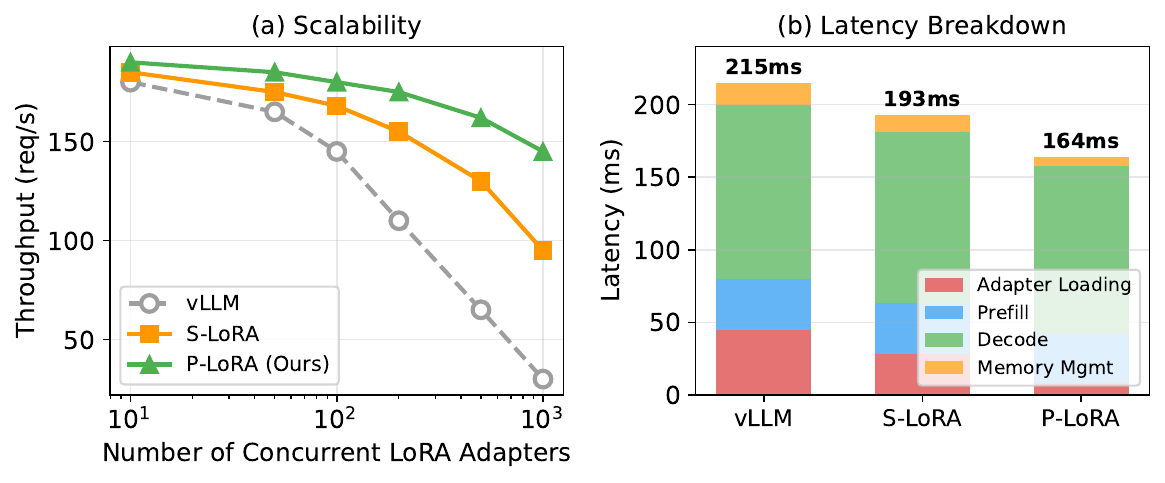}
\caption{(a) Throughput versus number of concurrent adapters. P-LoRA maintains higher throughput as adapter count increases. (b) Latency breakdown showing the contribution of each component. P-LoRA reduces adapter loading time by 82\% compared to vLLM.}
\label{fig:throughput}
\end{figure*}

Figure~\ref{fig:throughput} presents throughput scalability and latency breakdown results.

As the number of concurrent adapters increases, vLLM's throughput degrades rapidly due to frequent adapter merging/unmerging. S-LoRA maintains better scalability through batched computation but still suffers from reactive loading overhead. P-LoRA achieves 145 req/s with 1000 adapters, 1.52$\times$ higher than S-LoRA's 95 req/s.

The latency breakdown reveals that P-LoRA reduces adapter loading time from 45ms (vLLM) to 8ms through prefetching. Memory management overhead is also reduced from 15ms to 6ms due to page-based allocation eliminating fragmentation-related delays.

\subsection{Memory Efficiency}

\begin{figure*}[t]
\centering
\includegraphics[width=0.95\textwidth]{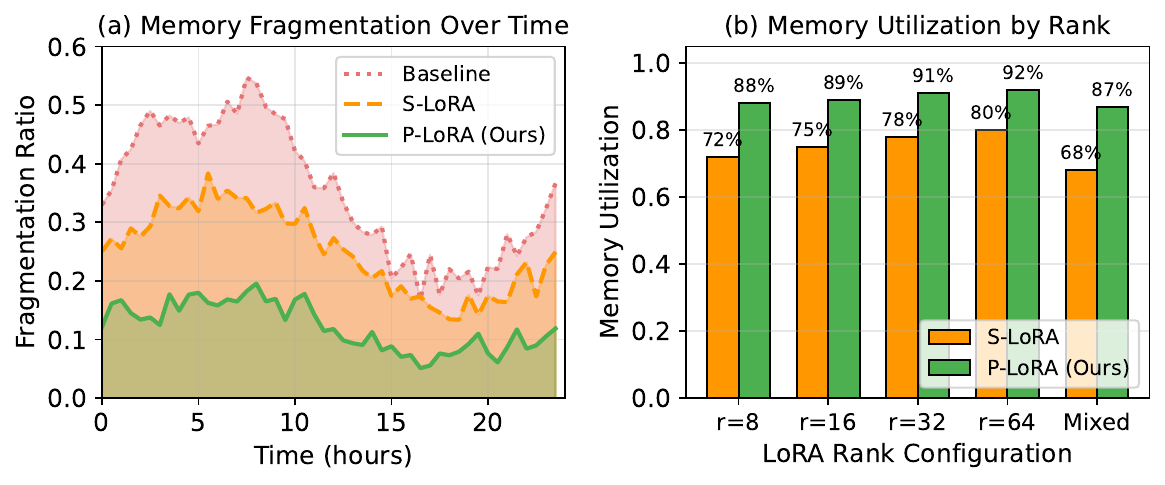}
\caption{(a) Memory fragmentation ratio over a 24-hour period. P-LoRA maintains consistently low fragmentation through page-based allocation. (b) Memory utilization across different adapter rank configurations.}
\label{fig:memory_efficiency}
\end{figure*}

Figure~\ref{fig:memory_efficiency} evaluates memory fragmentation and utilization.

P-LoRA maintains an average fragmentation ratio of 12\%, compared to 35\% for the baseline block allocator and 25\% for S-LoRA's unified paging. The improvement is most pronounced during high-churn periods when adapters are frequently swapped.

Memory utilization remains above 87\% across all rank configurations with P-LoRA. Notably, mixed-rank workloads show the largest improvement: 87\% utilization versus 68\% for S-LoRA. This is because page-based allocation handles heterogeneous sizes uniformly.

\subsection{Latency Under Varying Load}

\begin{figure*}[t]
\centering
\includegraphics[width=0.95\textwidth]{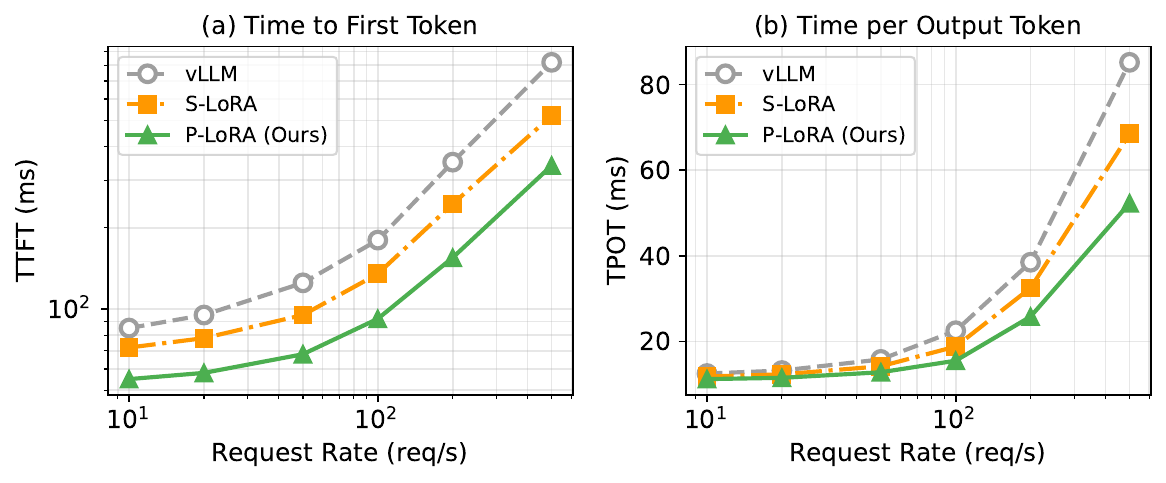}
\caption{(a) Time-To-First-Token (TTFT) versus request rate. (b) Time-Per-Output-Token (TPOT) versus request rate. P-LoRA maintains lower latency across all load levels. Note: axes use log scale.}
\label{fig:latency_workload}
\end{figure*}

Figure~\ref{fig:latency_workload} shows TTFT and TPOT under different request rates. Note that the axes use logarithmic scale to better visualize the wide range of values.

At 500 req/s, P-LoRA achieves average TTFT of 340ms compared to 520ms for S-LoRA (35\% reduction) and 820ms for vLLM (59\% reduction). TPOT shows similar trends, with P-LoRA maintaining 52.3ms versus 68.5ms for S-LoRA at maximum load.

\subsection{Ablation Study}

\begin{figure*}[t]
\centering
\includegraphics[width=0.95\textwidth]{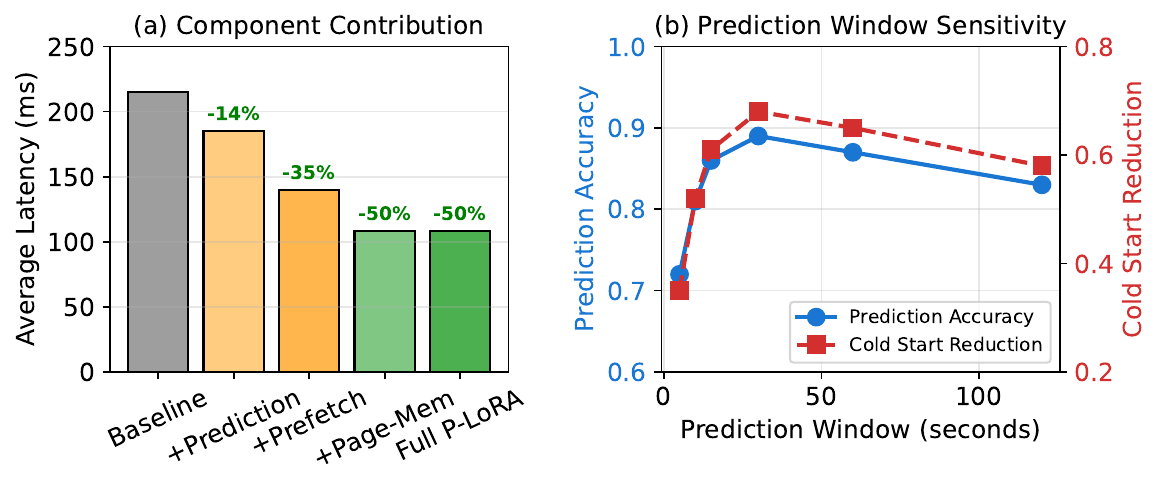}
\caption{(a) Contribution of each P-LoRA component to latency reduction compared to a vLLM-based baseline. (b) Sensitivity to prediction window size showing trade-off between prediction accuracy and cold start reduction.}
\label{fig:ablation}
\end{figure*}

Figure~\ref{fig:ablation} presents ablation results decomposing the contribution of each component. The baseline represents a vLLM-based system without P-LoRA's optimizations.

Adding prediction alone reduces latency by 14\%, prefetching adds another 21\%, and page-based memory management contributes an additional 15\%. The page-based allocation reduces latency by eliminating fragmentation-related memory compaction and eviction delays. The full system achieves 50\% latency reduction compared to the baseline.

The prediction window analysis reveals an optimal window of 30 seconds, achieving peak prediction accuracy of 89\% (compared to the 86\% overall average) and cold start reduction of 68\%. Shorter windows lack sufficient history, while longer windows include stale patterns that reduce accuracy.

\subsection{Overhead Analysis}

Table~\ref{tab:overhead} summarizes the overhead of P-LoRA's additional components.

\begin{table}[t]
\centering
\caption{Overhead of P-LoRA Components}
\label{tab:overhead}
\begin{tabular}{lcc}
\toprule
\textbf{Component} & \textbf{Latency (ms)} & \textbf{Memory (MB)} \\
\midrule
LSTM Predictor & 2.3 & 18 \\
Page Table & 0.4 & 32 \\
Prefetch Scheduler & 0.8 & 8 \\
\midrule
\textbf{Total} & \textbf{3.5} & \textbf{58} \\
\bottomrule
\end{tabular}
\end{table}

The total overhead of 3.5ms and 58MB is negligible compared to the improvements achieved. The LSTM predictor runs on CPU cores that would otherwise be idle during GPU inference.

\section{Related Work}
\label{sec:related}

\textbf{LoRA Serving Systems.} S-LoRA~\cite{sheng2024slora} introduced unified paging for concurrent LoRA serving. Punica~\cite{chen2023punica} developed SGMV kernels for batched LoRA computation. dLoRA~\cite{wu2024dlora} proposed dynamic adapter orchestration. These systems focus on efficient computation but employ reactive loading strategies.

\textbf{Serverless LLM Inference.} ServerlessLLM~\cite{fu2024serverlessllm} addresses cold start through multi-tier checkpoint loading. FlashAttention~\cite{dao2022flashattention} optimizes memory access patterns for attention computation. Our work is complementary, focusing on adapter-level optimization.

\textbf{Workload Prediction.} Time-series forecasting for cloud workloads has been extensively studied~\cite{ullah2023intelligent,bi2021integrated}. We adapt these techniques specifically for adapter access prediction in multi-LoRA serving scenarios.

\textbf{Memory Management.} GMLake~\cite{guo2024gmlake} proposed virtual memory stitching for DNN training. PagedAttention~\cite{kwon2023vllm} applies paging to KV cache management. We extend paging concepts to adapter weight management with distinct requirements.

\section{Discussion}
\label{sec:discussion}

While P-LoRA demonstrates improvements in throughput and latency, several limitations warrant discussion. The prediction accuracy depends on workload regularity; highly irregular or adversarial patterns may reduce prefetching benefits. The page-based approach introduces minor overhead for address translation, though this is amortized over batch processing. Future work could explore adaptive page sizes based on workload characteristics and integration with other memory optimization techniques.

\section{Conclusion}
\label{sec:conclusion}

We presented P-LoRA, a proactive and fragmentation-aware serverless inference system for LoRA-based LLMs. By introducing LSTM-based traffic prediction and page-based adapter memory management, P-LoRA achieves 1.52$\times$ throughput improvement over S-LoRA with 35\% reduction in average TTFT. The prediction component achieves 86\% average accuracy (up to 89\% under optimal settings) with minimal overhead, while page-based allocation maintains memory utilization above 87\% across heterogeneous adapter configurations. These results suggest that combining prediction-based proactive management with fine-grained memory allocation offers a promising direction for efficient serverless multi-LoRA inference.

\bibliographystyle{IEEEtran}
\bibliography{references}

\end{document}